\documentstyle[11pt,epsfig]{article}
\begin{document}
\title{Semiclassical Distorted Wave Model Analysis of Backward
Proton Emission from $(p,p^{\prime}x)$ Reactions at Intermediate
Energies}
\author{M.K. Gaidarov$^{a,b}$, Y. Watanabe$^a$, K. Ogata$^c$, \\
        M. Kohno$^d$, M. Kawai$^c$, A.N. Antonov$^b$}
\date{$^a${\it Department of Advanced Energy Engineering Science,\\
               Kyushu University, Kasuga, Fukuoka 816-8580, Japan}\\
$^b${\it Institute of Nuclear Research and Nuclear Energy,\\
         Bulgarian Academy of Sciences, Sofia 1784, Bulgaria}\\
$^c${\it Department of Physics, Kyushu University,\\
         Fukuoka 812-8581, Japan}\\
$^d${\it Physics Division, Kyushu Dental College,\\
         Kitakyushu 803-8580, Japan}}
\maketitle

A semiclassical distorted wave (SCDW) model with Wigner transform
of one-body density matrix is presented for multistep direct
$(p,p^{\prime}x)$ reactions to the continuum. The model uses
Wigner distribution functions obtained in methods which include
nucleon-nucleon correlations to a different extent, as well as
Woods-Saxon (WS) single-particle wave function. The higher
momentum components of target nucleons that play a crucial role in
reproducing the high-energy part of the backward proton spectra
are properly taken into account. This SCDW model is applied to
analyses of multistep direct processes in
$^{12}$C$(p,p^{\prime}x)$, $^{40}$Ca$(p,p^{\prime}x)$ and
$^{90}$Zr$(p,p^{\prime}x)$ in the incident energy range of
150--392 MeV. The double differential cross sections are
calculated up to three-step processes. The calculated angular
distributions are in good agreement with the experimental data, in
particular at backward angles where the previous SCDW calculations
with the WS single-particle wave function showed large
underestimation. It is found that the result with the Wigner
distribution function based on the coherent density fluctuation
model provides overall better agreement with the experimental data
over the whole emission energies.

\section{Introduction}

Preequilibrium processes in nuclear reactions at intermediate
energies are known to be dominated by multistep direct (MSD)
processes. Among the models proposed in the years, the
semiclassical distorted wave model \cite{Luo90,Kawai92,Wata93} has
proved its efficiency to describe MSD, especially for reactions at
intermediate energies. It is based on the DWBA expansion of
$T$-matrix elements and the cross section formula has no free
adjustable parameter allowing a simple intuitive interpretation.
The SCDW model has been applied to analyses of inclusive cross
sections and spin observables in $(p,p^{\prime}x)$ and $(p,nx)$
reactions over incident energy region of 60 to 400 MeV and wide
range of mass number
\cite{Wata99,Weili99,OgataMed,Ogata99,Ogata2002}. Recently,
several extensions and improvements of the SCDW model have been
made. They include calculations up to three-step processes
\cite{Wata99}, incorporation of a single-particle (s.p.) wave
function of target nuclei by using the Wigner transform of a
one-body density matrix (OBDM) \cite{Weili99}, introduction of
phenomenological effective mass $m^{*}$ of a nucleon in the target
nucleus \cite{OgataMed} and analyses of spin observables
\cite{Ogata99,Ogata2002} which contain important information on an
effective interaction in nuclear medium. As a result, SCDW model
calculations can well reproduce the experimental data of double
differential inclusive cross sections (DDX's) for
$(p,p^{\prime}x)$ and $(p,nx)$ reactions. The model, however,
underestimated the cross sections at very backward angles. This
suggests that improvement of the SCDW model might be achieved by
the inclusion of the high momentum components of the target
nucleons.

The problem of backward proton production in proton-nucleus
collisions has been a subject of much interest for several years.
The main difficulty arises from the fact that the reaction
involves a very large momentum transfer with only a small energy
transfer. One of the models which have been applied to solve the
problem is based on the correlated two-nucleon-cluster picture
\cite{Hane86}. A good description of the energy spectra even at
the very high energy backward protons is achieved with this model.
The main reason is that a phenomenological momentum distribution
with high momentum components was employed \cite{Fuji86}. These
components were found to be quite sensitive to the problem of
backward proton production and were considered to be an effect due
to many-nucleon correlations. Within the model \cite{Hane86}
single scattering and two-nucleon cluster mechanisms at backward
angles have been considered. Although there are some similarities
between these contributions and corresponding one- and two-step
processes, the advantage of the SCDW model is that the
contribution from three-step process can be also obtained.
Moreover, in SCDW model the distortion effects for the projectile
and observed protons are carefully considered within DWBA while
the calculated single scattering mechanism in \cite{Hane86} has
evaluated the distortion effect approximately by the Glauber
theory.

In Ref. \cite{Weili99} the SCDW model was modified so that
realistic s.p. wave functions in a finite range potential can be
used in terms of the Wigner transform of OBDM. Single-particle
models with harmonic oscillator and Woods-Saxon potentials were
used to describe the nuclear states for $^{90}$Zr$(p,p^{\prime}x)$
reactions at 80 and 160 MeV instead of the local density Fermi gas
(LFG) model used in previous analyses \cite{Wata99,Ogata99}. It
was concluded in \cite{Weili99} that the SCDW model with realistic
s.p. wave functions gives larger one-step cross sections at
backward angles, which result in better agreement with the
experimental data over a wider angular range than the model with
LFG. The calculations still somewhat underestimated the cross
section at very large angles. Therefore, apart from the necessity
of involving contributions of higher MSD processes the inclusion
of higher momentum components of target nucleons becomes apparent.
In the present paper we extend the SCDW model in terms of the
Wigner transform of OBDM by considering the high-momentum
components of nucleon momentum distributions deduced from several
theoretical methods. This extended SCDW model is applied to
$^{12}$C$(p,p^{\prime}x)$ reaction at 150 MeV
\cite{Steyn2001,Fortsch88}, $^{40}$Ca$(p,p^{\prime}x)$ reaction at
186 \cite{Steyn2000} and 392 MeV \cite{Cowley2000} and
$^{90}$Zr$(p,p^{\prime}x)$ at 160 MeV \cite{Richter94} and
comparison with the experimental data is made.

The aim of our work is to analyze the angular distributions of the
backward scattered protons through testing of different models for
the nuclear states which account for nucleon-nucleon (NN)
correlations. Such a systematic study could lead to a further
refinement of the SCDW model and to a better agreement of the DDX
with the experimental data at backward angles due to a proper
inclusion of high-momentum components of the nuclear Fermi motion.

The paper is organized as follows. A brief formulation of the SCDW
model and the proposed WT within different theoretical models are
given in Sections II and III, respectively. The results of the
calculations are presented and discussed in Section IV. The
summary of the present work is given in Section V.

\section{Outline of the SCDW model}

The formulation of the SCDW model used in the present analysis has
been described in details elsewhere
\cite{Weili99,Ogata99,Ogata2002}, therefore only the outline is
mentioned here.

Let us consider the one-step process in $(N,N^{\prime}x)$
reactions in which a target nucleon is excited from an initial
single particle state $\alpha $ with the energy
$\varepsilon_{\alpha }$ to a final one $\beta $ in the continuum
with the energy $\varepsilon_{\beta }$. A final expression of the
inclusive $(N,N^{\prime}x)$ double differential cross section for
the one-step process is given by
\begin{equation}
     \frac{\partial ^2\sigma ^{(1)}} {\partial E_f\partial \Omega _f}
  =
    \left ({A}\over{A+1} \right)^2
    \int d{\bf r}
    \frac{k_f/k_f({\bf r})}{k_i/k_i({\bf r})}
    |\chi^{(+)}_i({\bf r})|^2|\chi^{(-)}_f({\bf r})|^2 \\
   \sum_{\tau} {\left( {{{\partial ^2\sigma } \over {\partial E_f\partial
    \Omega _f}}}\right)}^{\tau}_{\bf r} \rho_{\tau}({\bf
    r}),
    \label{x-prev-new}
\end{equation}
where $A$ is the target mass number, $k_{c}$ and $k_{c}({\bf r})$
($c$=$i$ or $f$) the wave number at infinity and the local wave
number in the initial ($i$) and final ($f$) channels,
$\chi_{i}^{+}$ ($\chi_{f}^{-}$) the distorted wave in the initial
(final) channel, and $\rho_{\tau}({\bf r})$ ($\tau $=$p$ or $n$)
the nucleon density for proton ($p$) or neutron ($n$) as the
struck nucleon.

The local average nucleon-nucleon (NN) scattering cross section
for collision between a leading particle, $N$, and a target
nucleon, $\tau $, is given by
\begin{eqnarray}
    {\left( {\partial ^2\sigma } \over {\partial E_f\partial
    \Omega _f} \right)}^{\tau}_{\bf r}
 & = &
 \frac{2}{(2\pi)^3}
         \frac{1}{\rho_{\tau}({\bf r})}
         \frac{k_f({\bf r})}{k_i({\bf r})}
         \int \int d{\bf k}_{\alpha} d{\bf k}_{\beta}
         f_h^{\tau}({\bf k}_\alpha,{\bf r})
         [2-f_h^{\tau}({\bf k}_\beta ,{\bf r})] \nonumber\\
  & \times & \left ({{\partial \sigma}\over{\partial \Omega}} \right)_{N\tau}
         \delta( {\bf k}_f({\bf r})
                +{\bf k}_\beta
                -{\bf k}_i({\bf r})
                -{\bf k}_\alpha   )
         \delta (\varepsilon_\beta -
                 \varepsilon_\alpha -\omega ),
\label{localNN-pn}
\end{eqnarray}
where ${\bf k}_{\alpha }$(${\bf k}_{\beta }$) is the nucleon
momentum for a single-particle state with the energy
$\varepsilon_{\alpha }$($\varepsilon_{\beta }$), $\omega $ the
energy transfer, ($\partial \sigma /\partial \Omega $)$_{N\tau}$
the two-nucleon scattering cross section and $f_{h}^{\tau }({\bf
k},{\bf r})$ is the Wigner transform of the OBDM for the hole
states of proton and neutron. In Eq. (\ref{localNN-pn}) the sum
rule of the WT for the hole and particle states is used:
\begin{equation}
f_{h}({\bf k},{\bf r})+f_{p}({\bf k},{\bf r})=2,
\end{equation}
where $f_{p}({\bf k},{\bf r})$ stands for the WT for the particle
states. The factor $f_h^{\tau}({\bf k}_\alpha,{\bf
r})[2-f_h^{\tau}({\bf k}_\beta ,{\bf r})]$ in Eq.
(\ref{localNN-pn}) represents the "probability" of the momentum
state $\hbar {\bf k}_{\alpha }$ being occupied and $\hbar {\bf
k}_{\beta }$ being unoccupied. The Pauli blocking effect is
reflected in this factor. The difference between the
single-particle energies in the second $\delta$-function in the
right-hand side of Eq. (\ref{localNN-pn}) is given by
\begin{equation}
\varepsilon_{\alpha}-\varepsilon_{\beta}=\frac{\hbar^{2}}{2m^{*}({\bf
r})} [k_{\beta}^{2}({\bf r})-k_{\alpha}^{2}({\bf r})],
\label{eq:effmass}
\end{equation}
using the effective mass of a nucleon, $m^{*}({\bf r})$
\cite{Mahaux88}.

The extension to higher-step processes is straightforward and the
same final expressions as in Ref. \cite{Wata99} with the local
average NN scattering cross sections given by Eq.
(\ref{localNN-pn}) for successive collision points are also
deduced for two- and three-step processes.

\section{Wigner transform of one-body density matrix using
different theoretical models}

We introduce formulae for the Wigner transform (WT) derived in
some theoretical methods, namely calculated with WS s.p. wave
function (denoted as WS), with natural orbitals (NO) obtained from
the Jastrow correlation method (JCM) which accounts for
short-range correlations (SRC) in the case of $^{40}$Ca (denoted
as JCM), from the approach based on the local density
approximation (LDA) including the effects of SRC (denoted as
LDA+SRC) and from the coherent density fluctuation model (denoted
as CDFM).

\subsection{WT with WS}
For a single-particle potential, for instance Woods-Saxon
potential, the WT for the hole states of proton or neutron is
given by \cite{Weili99}
\begin{equation}
   f_{h}^{\tau}({\bf k},{\bf r})
  = \sum_{nlj}\frac{2j+1}{2l+1}
  \int_{0}^{\infty}d{\bf s}
      e^{-i{\bf k}{\cdot}{\bf s}}
      \sum_{m}
      \phi_{nlmj}^{\tau}({\bf r} + {\bf s}/2)
      \phi_{nlmj}^{\tau *}({\bf r} - {\bf s}/2),
\label{eq:wswigner}
\end{equation}
where $\phi_{nlmj}^{\tau}({\bf r})$ is the s.p. wave function for
the target nucleon and the sum runs over all the occupied orbits
$nlj$ of protons or neutrons.

\subsection{WT with JCM}
A model independent way to define a set of single-particle wave
functions and occupation probabilities uniquely from the
correlated OBDM $\rho({\bf r},{\bf r^{\prime}})$ is to use its
natural orbital representation (NOR) \cite{Jas55}
\begin{equation}
\rho_{NOR}({\bf r},{\bf r^{\prime}})=\sum_{\alpha }n_{\alpha}
\psi_{\alpha}^{*}({\bf r}) \psi_{\alpha}({\bf r^{\prime}}).
\label{eq:norobdm}
\end{equation}
The normalized eigenfunctions $\psi_{\alpha}({\bf r})$ of
$\rho({\bf r},{\bf r^{\prime}})$, the so-called natural orbitals,
form a complete orthonormal set. The associated eigenvalues
$n_{\alpha}$ called natural occupation numbers define the
probability ($0\leq n_{\alpha}\leq 1$) of the natural orbital
$\psi_{\alpha}({\bf r})$ occupation in the ground state $\Psi$.
Within the NOR the Wigner transfrm of $\rho({\bf r},{\bf
r^{\prime}})$ can be derived as follows:
\begin{equation}
f({\bf k}_{c},{\bf r})=\sum_{c}n_{c}h_{c}^{NO}({\bf k}_{c},{\bf
r}) = \sum_{c}n_{c}\int_{0}^{\infty}d{\bf s}e^{-i{\bf k}_{c}{\bf
s}}\psi_{c}({\bf r}+{\bf s}/2) \psi_{c}^{*}({\bf r}-{\bf s}/2)
\label{eq:nowigner}
\end{equation}
We would like to note that the summation in Eq.
(\ref{eq:nowigner}) is taken over all hole- as well as
particle-state NO included in the calculation of the Wigner
transform. In the present work we use the results for the natural
orbitals and for the occupation numbers in $^{40}$Ca obtained in
Ref. \cite{Sto93} within the JCM in its low-order approximation.
As it has been already done in \cite{Weili99} for the s.p. wave
functions, in the present work each NO $\psi_{c}({\bf r})$ is
expanded in terms of the Gaussian-type basis function
\begin{equation}
\psi_{c}({\bf r})=\sum_{\upsilon=1}^{N}a_\upsilon^{(c)}
\exp(-\kappa_\upsilon^2{\,}r^2), \label{eq:gauss}
\end{equation}
where $N$ is the number of the basis functions, $\kappa_\upsilon$
is given by a geometrical progression \cite{Kami88},
$\kappa_\upsilon=\kappa_1(\kappa_N/\kappa_1)^{(\upsilon-1)/(N-1)}$.
The $\{N,\kappa_1,\kappa_N\}$ are the input parameters. The
expansion coefficients $a_\upsilon^{(c)}$ are determined by linear
fitting of the basis functions (\ref{eq:gauss}) to the numerical
data for the natural orbitals. Then each partial Wigner transform
is calculated in accordance with Eq. (2.39) in Ref. \cite{Weili99}
and after applying Eq. (\ref{eq:nowigner}) the total Wigner
transform for the NOR of $\rho ({\bf r},{\bf r^{\prime} })$ is
obtained.

\subsection{WT with LDA+SRC}
It is well known that the inclusion of correlations in nuclear
matter modifies the occupation probability predicted by the local
density Fermi gas model. According to Ref. \cite{Strin90} one can
introduce Wigner transform in a general way:
\begin{equation}
f(k_{F}({\bf r}),k)=\Theta(k_{F}({\bf r})-k)+\delta f(k_{F}({\bf
r}),k),
\label{eq:ldawigner}
\end{equation}
where $\Theta(k_{F}-k)$ corresponds to the WT of the LFG model,
while $\delta f(k_{F},k)$ is entirely due to the effects of
dynamical correlations induced by the NN interaction. The local
Fermi momentum $k_{F}(r)$ is related to the mass density through
the relation
\begin{equation}
k_{F}(r)=\left [\frac{3}{2}\pi^{2}\rho(r)\right ]^{1/3}.
\label{eq:kf}
\end{equation}
By definition of $k_{F}(r)$ one has $\int \delta f(k_{F}({\bf
r}),k)d{\bf k}=0$. It was shown in \cite{Strin90} that for a
finite nucleus the separation between mean-field contribution and
correlation effects can be performed in an analogous way. For
convenience more phenomenological procedure based on the results
of the lowest order cluster (LOC) approximation developed in
\cite{Flynn84} has been followed to evaluate explicitly the
correlated term. Choosing a correlation function of the form
\begin{equation}
f(r)=1-e^{-\beta^{2}r^{2}},
\label{eq:corrfunc}
\end{equation}
the LOC gives for $\delta f(k_{F},k)$
\begin{equation}
\delta f(k_{F}({\bf r}),k)=\left [Y(k,8)-k_{dir}\right
]\Theta(k_{F}({\bf r})-k)+8\left
\{k_{dir}Y(k,2)-[Y(k,4)]^{2}\right \}, \label{eq:corrterm}
\end{equation}
where
\begin{equation}
c_{\mu}^{-1}Y(k,\mu)=\frac{e^{-\tilde{k}_{+}^{2}}-e^{-\tilde{k}_{-}^{2}}}{2\tilde{k}}+
\int_{0}^{\tilde{k}_{+}}e^{-y^{2}}dy+sgn
(\tilde{k}_{-})\int_{0}^{|\tilde{k}_{-}|}e^{-y^{2}}dy
\label{eq:ygrek}
\end{equation}
with
\begin{equation}
c_{\mu}=\frac{1}{8\sqrt{\pi}}\left(\frac{\mu}{2}\right
)^{3/2},\;\;\; \tilde{k}=\frac{k}{\beta\sqrt{\mu}},\;\;\;
\tilde{k}_{\pm}=\frac{k_{F}\pm k}{\beta\sqrt{\mu}},\;\;\;
sgn(x)=\frac{x}{|x|}.
\label{eq:constants}
\end{equation}
The quantity
\begin{equation}
k_{dir}=\frac{2k_{F}^{3}}{3\pi^{2}}\int(f(r)-1)^{2}d{\bf
r}=\frac{1}{3\sqrt{2\pi}}\left(\frac{k_{F}}{\beta}\right )^{3}
\label{eq:jaspar}
\end{equation}
is the direct part of the Jastrow wound parameter.

\subsection{WT with CDFM}
The CDFM has been suggested in \cite{Ant79,Ant88,Ant93} as a model
for studying characteristics of nuclear structure and nuclear
reactions based on the local density distribution as a variable of
the theory and using the essential results of the infinite nuclear
matter theory. The model is introduced using the main ansatz of
the generator coordinate method for the many-body function and the
delta-function approximation for the overlap and energy kernels of
the corresponding integral equation for the weight function. In
the CDFM the Wigner distribution function can be written in the
form:
\begin{equation}
f({\bf r},{\bf k})=\int_{0}^{\infty}|f(x)|^{2}\Theta(x-|{\bf
r}|)\Theta(k_{F}(x)-|{\bf k}|)dx,
\label{eq:cdfmwigner}
\end{equation}
where the weight function $f(x)$ in the generator coordinate
method is determined under the condition
\begin{equation}
\int_{0}^{\infty} |f(x)|^{2}dx=1.
\end{equation}
In the case of monotonically-decreasing density distributions
$(d\rho /dr\leq 0)$ one can obtain a relation of the weight
function $f(x)$ with the density distribution:
\begin{equation}
|f(x)|^{2}=-\frac{1}{\rho_{0}(x)} \left. \frac{d\rho(r)}{dr}\right
|_{r=x},
\label{eq:weightf}
\end{equation}
where
\begin{equation}
\rho_{0}(x)=3A/4\pi x^{3}
\end{equation}
and the generator coordinate $x$ is the radius of a sphere
containing all $A$ nucleons uniformly distributed in it (the
so-called "flucton").

\section{Results of calculations and discussion}

The input data used in the calculations of Wigner transforms can
be summarized as follows. The single-particle wave function in Eq.
(\ref{eq:wswigner}) were calculated using a Woods-Saxon potential
with the radius parameter $r_{0}$=1.27 fm and the surface
diffuseness parameter $a$=0.67 fm, including the isovector term
and the Coulomb potential. The set
$\{N,\kappa_1,\kappa_N\}$=$\{20,1,0.1\}$ is taken for the WS
potential and for the WT with NOR. For the correlation factor in
Eq. (\ref{eq:corrfunc}) we adopt the same value $\beta$=1.1 fm
$^{-1}$ as in Ref. \cite{Strin90} which is taken from the
microscopic nuclear matter calculations \cite{Fant84} but
reproduces also well the results for the nucleon momentum
distribution $n(k)$ of finite nuclei. This value leads to
$k_{dir}$=0.23 [Eq. (\ref{eq:jaspar})]. In the CDFM method as well
as in the LDA approach with SRC we use the nucleon density
distribution $\rho(r)$ of the Woods-Saxon shape with Negele's
geometrical parameters \cite{Negel70}. The proton and the neutron
densities are assumed to be given by $(Z/A)\rho(r)$ and
$(N/A)\rho(r)$, respectively. The local Fermi momentum $k_{F}(r)$
was obtained from Eq. (\ref{eq:kf}) using the same density
distribution.

As for the other input data for the SCDW calculations, we use
basically the same ones as in Ref. \cite{Ogata2002}. The global
optical potentials based on Dirac phenomenology of Hama {\it et
al.} \cite{Hama90} for protons and Ishibashi {\it et al.}
\cite{Ishi97} for neutrons were used. For a simplicity of the
numerical calculations we neglect the spin-orbit coupling in the
distorting potentials. The nonlocality correction for distorted
waves and WS s.p. wave functions was taken into account using the
Perey factor with range 0.85 \cite{Perey62}. An effective NN
interaction in terms of $G$ matrix parameterized by the Melbourne
group \cite{Amos2000} was employed. The two-nucleon scattering
cross section in Eq. (\ref{localNN-pn}) is calculated using this
$G$ matrix. In addition to \cite{Ogata2002}, an effective mass of
a target nucleon $m^{*}$ is used. We assume a simple WS form for
the $r$-dependence of $m^{*}$. The latter at $r$=0 was chosen to
be $m^{*}=0.8m$ with a bare mass $m$.

\begin{figure}[ht]
\centering\epsfig{file=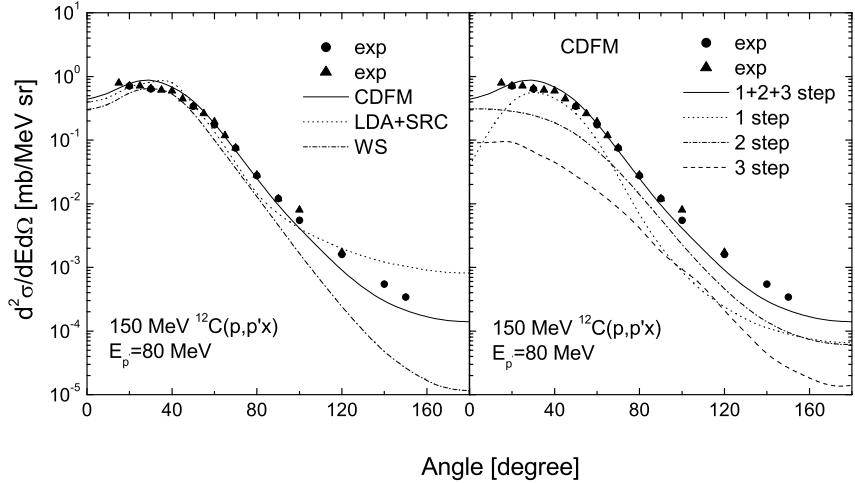,width=130mm}
\caption[]{Comparison of SCDW calculations with measured double
differential cross section for the reaction $^{12}$C($p,p'x$) at
150 MeV incident and 80 MeV emission energies (solid circles from
\cite{Steyn2001} and solid triangles from \cite{Fortsch88}). The
calculations with CDFM (solid line), LDA+SRC (dotted line) and WS
(dash-dotted) line are shown in the left panel. The CDFM
calculations of the cross sections of one- (dotted line), two-
(dash-dotted line) and three-step (dashed line) processes are
shown in the right panel, where the solid line corresponds to
their sum.} \label{cddx15080}
\end{figure}

Figure \ref{cddx15080} shows the SCDW double differential cross
sections with CDFM, LDA+SRC and WS for the
$^{12}$C$(p,p^{\prime}x)$ reaction at 150 MeV incident and 80 MeV
outgoing energies compared with two sets of experimental data
\cite{Steyn2001,Fortsch88}. A good overall agreement with the data
including backward angle region is obtained in the case of CDFM.
The SCDW cross section calculated with WS drops down at backward
angles, while those with LDA+SRC overestimates the experimental
data. As will be discussed later, the main reason for this
behavior is the different account for the high-momentum components
of the nuclear Fermi motion. In Fig. \ref{cddx15080} the
contributions of individual multistep processes when CDFM is used
for the same reaction are also plotted. One can see that proton
emission via one-step process is dominant in the angular region
from 20$^{\circ}$ to 50$^{\circ}$. Contributions of two- and
three-step processes become appreciable with increasing angles.
The role of higher-step processes has been already seen in
previous calculations \cite{Wata99, Weili99}.

\begin{figure}[ht]
\centering\epsfig{file=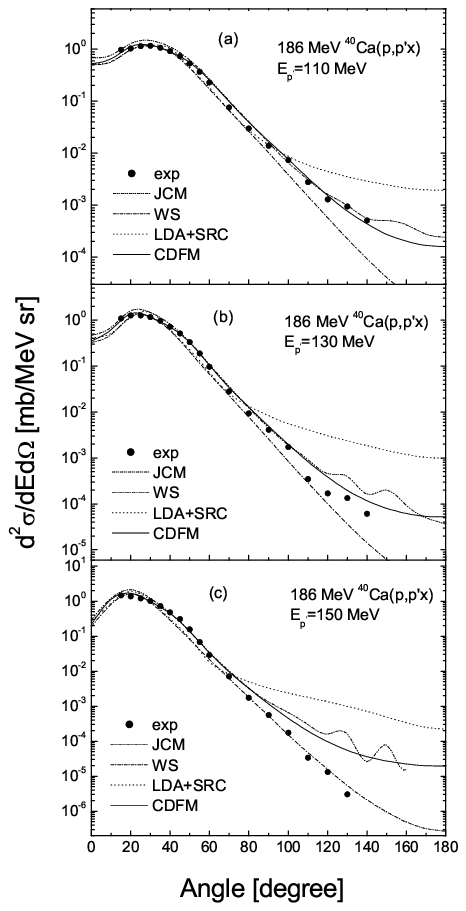,width=75mm}
\caption[]{Comparison of SCDW calculations with measured double
differential cross section for the reaction $^{40}$Ca($p,p'x$) at
186 MeV incident and 110 MeV (a), 130 MeV (b) and 150 MeV (c)
emission energies \cite{Steyn2000}. In each panel the calculations
with CDFM (solid line), JCM (dash-double dotted line), LDA+SRC
(dotted line) and WS (dash-dotted line) are shown.}
\label{caddx186}
\end{figure}

\begin{figure}[ht]
\centering\epsfig{file=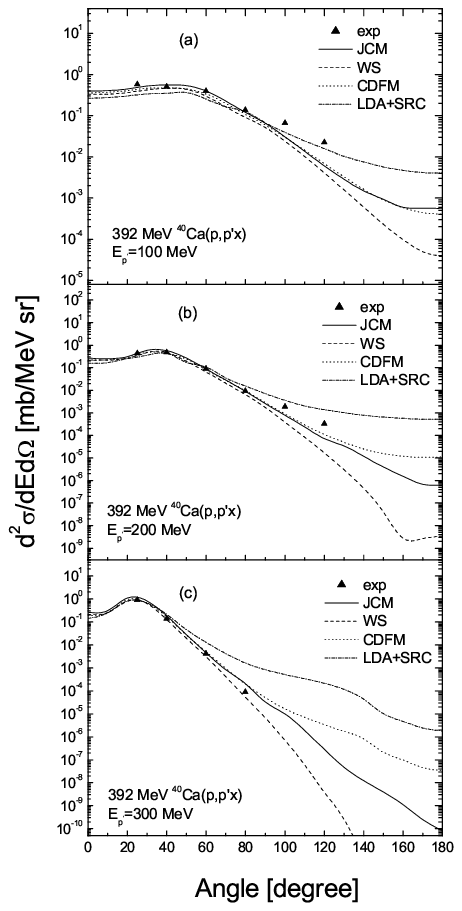,width=75mm}
\caption[]{Comparison of SCDW calculations with measured double
differential cross section for the reaction $^{40}$Ca($p,p'x$) at
392 MeV incident and 100 MeV (a), 200 MeV (b) and 300 MeV (c)
emission energies \cite{Cowley2000}. In each panel the
calculations with CDFM (dotted line), JCM (solid line), LDA+SRC
(dash-dotted line) and WS (dashed line) are shown.}
\label{caddx392}
\end{figure}

Comparisons between the cross sections calculated by using the
Wigner transforms obtained with all theoretical methods considered
and the measured cross sections for $(p,p^{\prime}x)$ on $^{40}$Ca
at 186 MeV and 392 MeV incident energies are given in Fig.
\ref{caddx186} and Fig. \ref{caddx392}, respectively. Both figures
confirm the fact that the cross sections at large angles depend
strongly on the theoretical model used to describe the nuclear
states. Fig. \ref{caddx186} shows a fair agreement of the CDFM and
JCM results with the data for emission energy of 110 MeV and to
less extent for energy of 130 MeV. At the same time, for emission
energy of 150 MeV the WS result is able to reproduce the angular
distribution at backward angles. Similar behavior can be seen from
Fig. \ref{caddx392}, where the double differential cross sections
calculated with CDFM and that with JCM are in a good agreement
with the experimental data. However, measurements over an extended
range of backward angles can test the various methods and, hence,
the role of NN correlations.

\begin{figure}[ht]
\centering\epsfig{file=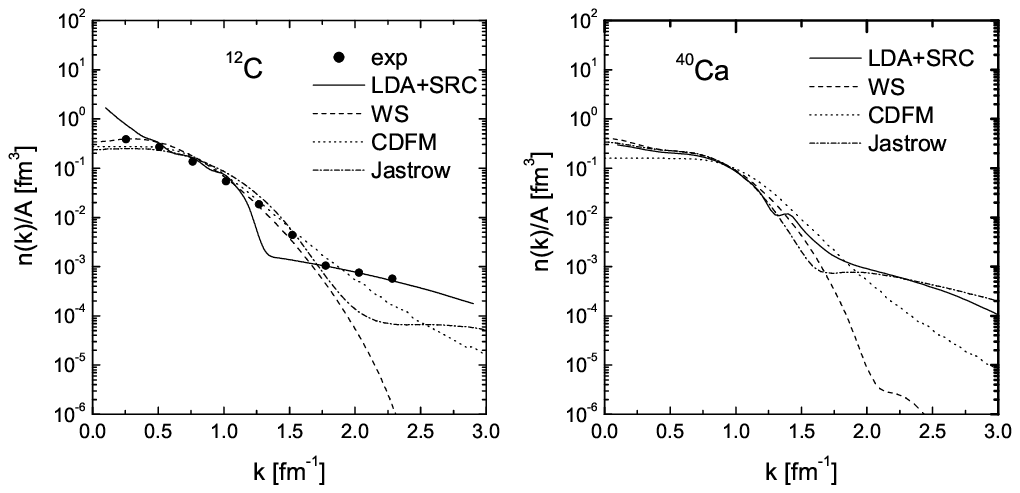,width=130mm} \caption[]{Comparison
of nucleon momentum distributions $n(k)$ (normalized to unity) of
$^{12}$C and $^{40}$Ca calculated with CDFM (dotted line), LDA+SRC
(solid line), WS (dashed line) and JCM \cite{Sto93} (dash-dotted
line). The experimental data for $n(k)$ of $^{12}$C are taken from
\cite{Cio91}.} \label{md}
\end{figure}

The improved description of the backward proton emission with the
SCDW model by the use of realistic models for nuclear states can
be explained by the difference in the momentum distribution of the
target nucleons. Figure \ref{md} shows the momentum distributions
$n(k)$ of $^{12}$C and $^{40}$Ca with CDFM, LDA+SRC, WS and JCM.
One sees that $n(k)$ calculated with all methods but not with the
WS contain high-momentum components at momenta larger than 1.5
fm$^{-1}$. The LDA+SRC \cite{Strin90} and Jastrow method
\cite{Sto93} account for short-range correlations, while the CDFM
\cite{Ant79}takes into account mostly long-range correlations
(LRC). The larger high-momentum component due to the stronger SRC
when using LDA+SRC causes the larger cross sections at backward
angles. Although $n(k)$ calculated in this method fits well the
experimental $y$-scaling data for $^{12}$C \cite{Cio91} at large
momenta, the corresponding cross sections overestimate the data.
Here we would like to note that the $n(k)$ of $^{40}$Ca obtained
after integration of the total WT $f({\bf k},{\bf r})$ with JCM
over ${\bf r}$ is not given in Fig. \ref{md} and it underestimates
the one obtained in \cite{Sto93} at $k\geq2$ fm$^{-1}$. This is
particularly due to the truncation of the particle-state NO's
included in the calculation of the WT of $^{40}$Ca. On the other
side, due to the lack of enough high-momentum components the
angular distributions calculated with WS underestimate the data at
backward angles. The nucleon momentum distributions obtained with
CDFM for both nuclei presented in Fig. \ref{md} also exhibit
high-momentum components, though smaller than the ones
corresponding to methods accounting for SRC. Fig. \ref{md} clearly
illustrates the strong dependence of backward proton production
from $(p,p^{\prime}x)$ reactions on the momentum distribution of
target nucleons.

\begin{figure}[ht]
\centering\epsfig{file=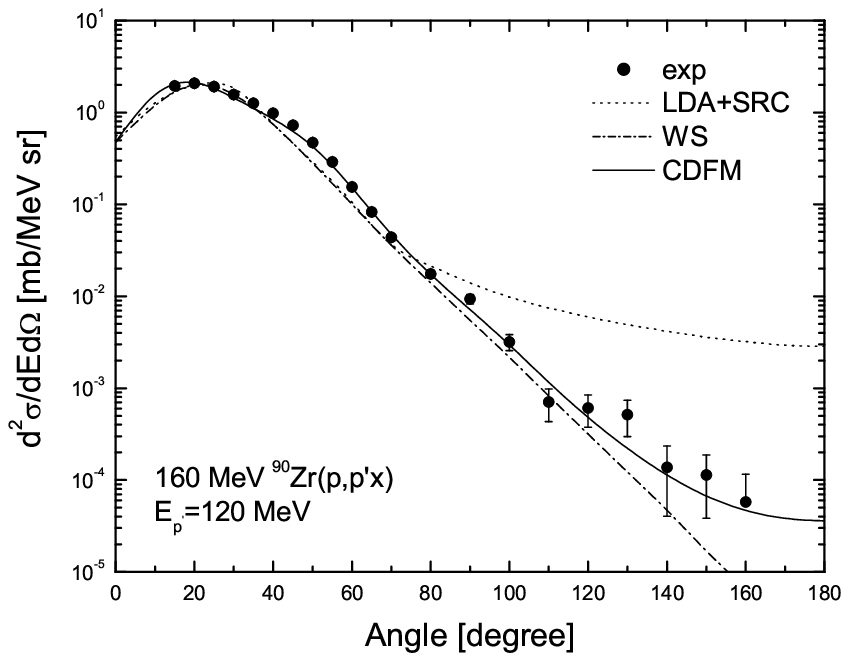,width=100mm}
\caption[]{Comparison of SCDW calculations with measured double
differential cross section for the reaction $^{90}$Zr($p,p'x$) at
160 MeV incident and 120 MeV emission energies \cite{Richter94}.
The calculations with CDFM (solid line), LDA+SRC (dotted line) and
WS (dash-dotted line) are shown.} \label{zrddx160120}
\end{figure}

\begin{figure}[ht]
\centering\epsfig{file=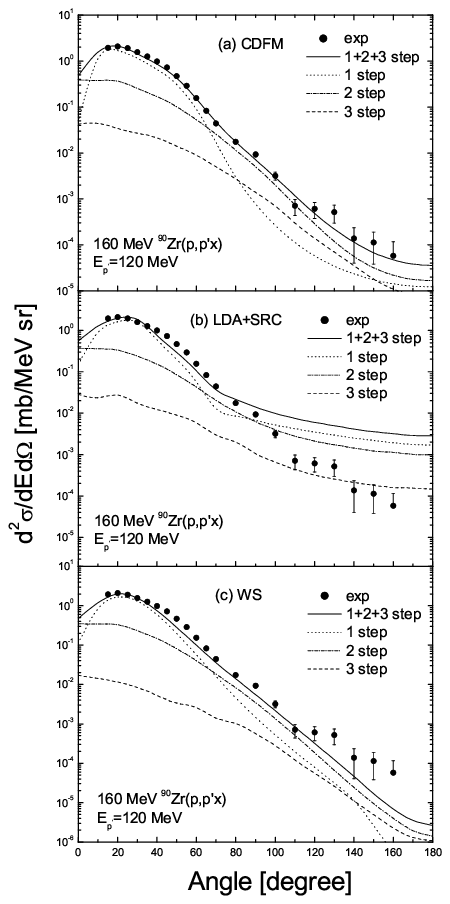,width=75mm}
\caption[]{Comparison of SCDW calculations with measured double
differential cross section for the reaction $^{90}$Zr($p,p'x$) at
160 MeV incident and 120 MeV emission energies \cite{Richter94}.
The calculations with CDFM, LDA+SRC and WS are shown in panels
(a), (b) and (c), respectively. In each panel the cross sections
of one- (dotted line), two- (dash-dotted line) and three-step
(dashed line) processes are shown. The solid line corresponds to
their sum.} \label{zrddxsteps}
\end{figure}

In Fig. \ref{zrddx160120} the SCDW cross sections for
$^{90}$Zr$(p,p^{\prime}x)$ at 160 MeV and emission energy of 120
MeV calculated with the CDFM, LDA+SRC and WS models are compared.
Again a very good agreement with the experimental data over the
forward, intermediate and backward angles is achieved when using
the Wigner transform obtained from CDFM. In Fig. \ref{zrddxsteps}
the contributions of individual multistep processes for the same
reaction and corresponding to the same models are plotted to show
their variation with the scattering angle. It is seen that for
this relatively large emission energy the one-step cross section
calculated with LDA+SRC exceeds the experimental cross section for
large angles. On the other hand, the one-step cross section
obtained with WS drops at angles less than 140$^{\circ}$
underestimating the experimental cross section at backward angles.

\begin{figure}[ht]
\centering\epsfig{file=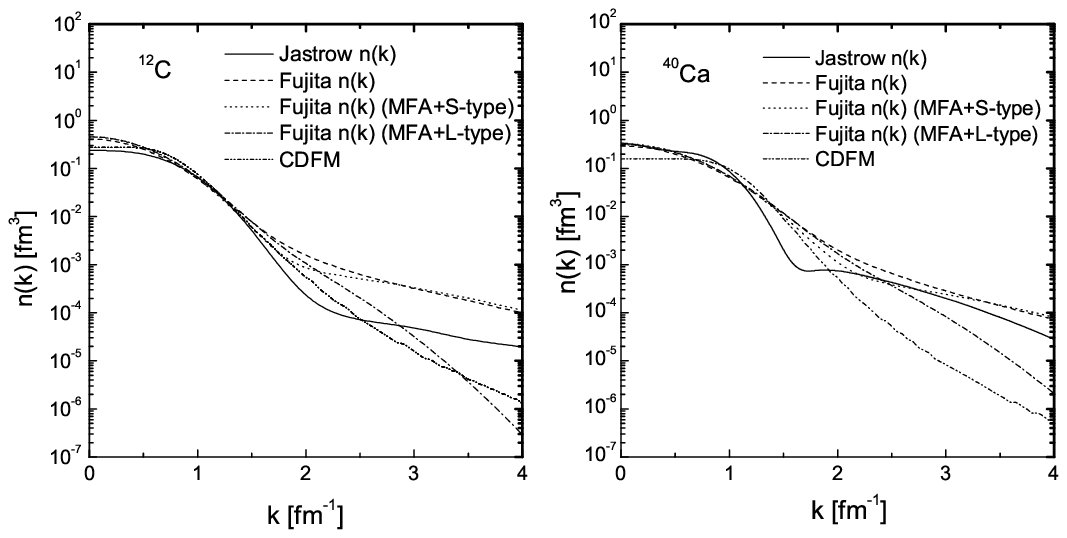,width=140mm}
\caption[]{Comparison of nucleon momentum distributions $n(k)$
(normalized to unity) of $^{12}$C and $^{40}$Ca calculated with
CDFM (dash-double dotted line), JCM \cite{Sto93} (solid line) and
by using Eq. (\ref{eq:fujita}): total $n(k)$ (dashed line),
MFA+S-type terms (dotted line) and MFA+L-type terms (dash-dotted
line).}
\label{fujimom}
\end{figure}

From the analyses of high energy heavy ion collision and deep
inelastic lepton-nucleus scattering, the following form of the
nucleon momentum distribution has been proposed \cite{Fuji86}:
\begin{equation}
n(k)=N[\exp(-k^{2}/p_{0}^{2})+\varepsilon_{0}\exp(-k^{2}/q_{0}^{2})+
\varepsilon_{1}\exp(-k^{2}/q_{1}^{2})],
\label{eq:fujita}
\end{equation}
where $N$ is a normalization factor and the values of the
parameters are determined to be $p_{0}=\sqrt{\frac{2}{5}}k_{F}$,
$\varepsilon_{0}\cong 0.03$, $q_{0}\cong \sqrt{3}p_{0}$,
$\varepsilon_{1}\cong 0.003$ and $q_{1}\cong 0.5$ GeV/c. The first
term of Eq. (\ref{eq:fujita}) corresponds to single-particle
contribution to $n(k)$ in the mean-field approximation (MFA). It
was noticed in \cite{Fuji86} that the second term in Eq.
(\ref{eq:fujita}) is due to the long-range correlations (L-type
high momentum component) while the third term was supposed to be
due to short-range correlations (S-type high momentum component).
Moreover, as it was also noticed in Ref. \cite{Fuji86} only the
high-momentum component generated by the Hartree-Fock correlation
should be included in the single scattering calculation. In order
to reveal better the role of different NN correlations for the
correct description of the backscattered proton spectra, we make
in Fig. \ref{fujimom} a comparison between the nucleon momentum
distributions calculated with CDFM and JCM and those calculated by
using Eq. (\ref{eq:fujita}) in the case of $^{12}$C and $^{40}$Ca
nuclei. Particularly, our interest is in the comparison of the
high-momentum components which the different methods contain. It
turns out from Fig. \ref{fujimom} that for $k>1.5$ fm$^{-1}$ the
curves corresponding to the inclusion only of the L-type term and
CDFM have similar behavior (especially for $^{12}$C), while the
results when including only the S-type term and Jastrow method are
close to each other. It is seen also from Fig. \ref{fujimom} that
the contribution of the S-type term dominates in the total $n(k)$
calculated by Eq. (\ref{eq:fujita}). Therefore, we suggest that
the long-range correlations play an important role to the problem
of backward proton production. This is confirmed by the good
agreement obtained in \cite{Hane86} and by our results obtained
with the use of Wigner transform with CDFM which lead to best
description of the experimental cross sections over wide target
muss number and emission energy regions.

\section{Summary}
The results of the present work can be summarized as follows:
\newline
i) Double-differential cross sections of
$^{12}$C$(p,p^{\prime}x)$, $^{40}$Ca$(p,p^{\prime}x)$ and
$^{90}$Zr$(p,p^{\prime}x)$ reactions are calculated, using SCDW
model and Wigner transform of the OBDM obtained with different
approaches, in the incident energy range of 150-392 MeV. A good
overall agreement with the experimental data is achieved including
the backward angles region. These analyses show that the present
SCDW model can describe successfully the MSD processes of
$(p,p^{\prime}x)$ reactions over the wide range of target mass
number and incident energies.
\newline
ii) It is found that the high-momentum components of the nucleon
momentum distribution are most responsible for describing the
angular distributions of the backscattered protons. Since CDFM
leads to the best agreement with the experimental cross sections
at wide emission energy region, it is demonstrated that the
long-range correlations which are related to the collective
nucleon motion are rather important than the short-range ones
generated in the other correlation methods.

A more detailed study on the backward proton emission in the
framework of the CDFM is in progress.

\section{Acknowledgments}
The authors wish to thank Professor A.A. Cowley for providing us
the experimental data from Ref. \cite{Steyn2000}. One of the
authors (MKG) is grateful to the JSPS's Postdoctoral Fellowships
for foreign researchers. This work was partly supported by a
Grant-in-Aid for Scientific Research of the Ministry of Education,
Science and Culture (No. 14002042).

\end{document}